# Controllable and Continuous Quantum Phase Transitions in Intrinsic Magnetic Topological Insulator

Shengjie Xu[1,†], Zhijian Shi[1,†], Ming Yang[2], Jingwei Zhang[1], Hang Xu[1], Haifeng Feng[1], Ningyan Cheng[3], Jianfeng Wang[1,*], Weichang Hao[1,2,*], Yi Du[1,2,*]

[1] *School of Physics, Beihang University, Haidian District, Beijing 100191, China*
[2] *Analysis & Testing Center of Beihang University. Beihang University, Beijing 100191, China*
[3] *Information Materials and Intelligent Sensing Laboratory of Anhui Province, Key Laboratory of Structure and Functional Regulation of Hybrid Materials of Ministry of Education, Institutes of Physical Science and Information Technology, Anhui University, Hefei, Anhui, China*

*†Shengjie Xu and Zhijian Shi contributed equally to this work.*
*\*Correspondence authors. E-mail: yi_du@buaa.edu.cn; whao@buaa.edu.cn; wangjf06@buaa.edu.cn*

**The intrinsic magnetic topological material MnBi$_2$Te$_4$ has demonstrated great potential to investigate the interplay between topology and magnetism, which opens up new avenues for manipulating non-trivial electronic states and designing quantum devices. However, challenges and controversies remain due to its inevitable *n*-type antisite defects, hindering the experimental realization of intrinsic magnetic topological phenomena and rendering the precise control of topological phase transitions (TPTs) unachievable. Here, we study a candidate material family, Mn$_{(1-x)}$Ge$_x$Bi$_2$Te$_4$, in which the heavy *n*-type doping features are strongly suppressed when the Ge content reaches 0.46, and multiple topological phases are well maintained with the surface Dirac point located near the Fermi level. Based on angle-resolved photoemission spectroscopy, transport measurements, and first-principles calculations, we reveal two magnetism-induced TPTs: the first is antiferromagnetic-ordering-induced transition from strong topological insulator to a magnetic topological insulator as revealed by gap opening of topological surface states; the second is external-magnetic-field-dependent transition from magnetic topological insulator to a Weyl semimetal with the gap reclosed. Our work paves the way for the realization of intrinsic magnetic topological states in MnBi$_2$Te$_4$ family and provides an ideal platform for achieving controllable and continuous TPTs towards future spintronic applications.**

## I. INTRODUCTION

Topological materials, classified by topological invariants and featuring robust boundary states, present a broad platform towards the development of dissipationless spintronics, information storage and quantum computation [1-4]. The introduction of magnetism into topological materials provides opportunities to investigate exotic phenomena arising from the interplay between magnetism and topology [5-7]. One successful example is the first realization of quantum anomalous Hall (QAH) effect in a Cr-doped (Bi, Sb)$_2$Te$_3$ film [8]. However, the weak magnetic exchange coupling and existing disorders in magnetically doped topological materials limit the observation of quantum behavior at extremely low temperatures and high magnetic fields, thus limiting their practical applications. Recent research efforts have concentrated on intrinsic magnetic topological materials due to their strong coupling between magnetism and topology, enabling the realization of topological phenomena at elevated temperatures [9,10]. Moreover, the attainment of intrinsic exotic magnetic topological states with nontrivial features near the Fermi level, as well as the effective control of their topological phase transitions (TPTs), are urgently needed for the subsequent design of innovative quantum devices.

As the first intrinsic magnetic topological insulator, MnBi$_2$Te$_4$ (MBT) has received extensive studies, which was predicted to host a variety of magnetic topological phases by tuning the magnetic order of MBT, including magnetic topological insulator (MTI) or Weyl semimetal (WSM) in three-dimensional (3D) bulk and Chern or axion insulator (AI) in two-dimensional (2D) films [11-16]. Remarkably, the QAH effect and AI state have been observed in thin films of MBT at relatively high temperatures [14,15]. However, the experimental observations in MBT encounter numerous challenges and controversies. The predicted ground state of MBT is an antiferromagnetic MTI with Dirac surface states opening a magnetic gap, which lays the foundation of QAH and AI states. While many angle-resolved photoemission spectroscopy (ARPES) measurements observed gapless surface states contradicting theoretical predictions [17,18]. Furthermore, the QAH effect in MBT was not widely observed. Except for the recent experimental realization of QAH effect in a five-layer film [14], most films did not observe quantized Hall resistance at zero fields [19-21]. All these controversies can be related to the existence of a large number of disorganized intrinsic Mn/Bi antisite defects (Bi$_{Mn}^+$ and Mn$_{Bi}^-$) in MBT as demonstrated by recent theoretical and experimental studies [22-25]. Consequently, since positive charge defects Bi$_{Mn}^+$ dominate in MBT, all the experimentally synthesized MBT samples are *n*-type doped, with their electronic properties dominated by trivial conduction bands (CBs). Actually, even the observed QAH effect in 2D films was realized under an extremely high gate voltage set by shifting the Fermi level to magnetic nontrivial gap [9,15]. In addition, recent studies have found that the Mn$_{Bi}^-$ antisite defect can strongly suppress the magnetic gap down to several meV, which may responsible for the gapless surface states observed in ARPES measurements [26]. Therefore, the effective control of antisite defects is one of the prerequisites to achieving high-temperature QAHE and controllable TPTs in MBT. Although various methods have been tried to improve the synthesis process, the unavoidable antisite defects still extremely difficult to eliminate due to the high migration barrier in kinetics [25]. Therefore, developing an effective way to regulate the antisite defects in MBT with its topologies and magnetism retained is crucial for the realization of intrinsic magnetic topological phenomena and controllable TPTs towards future applications [16].

In this work, with the aim of resolving the above issues, we adopted a Ge doping strategy in the Mn layer to regulate the ratio of Bi/Mn antisite defects. It was found that the antisite defects in Mn$_{(1-x)}$Ge$_x$Bi$_2$Te$_4$ (MGBT) can be regulated by adjusting the atomic ratio *x*, thereby tuning the position of the Fermi level. In the case of our Mn$_{0.54}$Ge$_{0.46}$Bi$_2$Te$_4$ sample, the Fermi level is situated nearly at the surface Dirac point, as revealed by ARPES measurements. Importantly, a unique magnetism dependent metal-insulator-metal transition on Mn$_{0.54}$Ge$_{0.46}$Bi$_2$Te$_4$ was confirmed by transport measurements, revealing a continuous magnetism induced TPTs process from a strong topological insulator (STI) with gapless surface states to an MTI with gapped surface states and then to a gapless WSM. All these electronic band structures are reproduced by our first-principles calculations. This work provides a very promising platform and a novel manipulating approach for studying the intrinsic magnetic topological phenomena and magnetism induced TPTs, which is of great significance for the development of spintronics and quantum computing.

## II. METHODS

Crystal growth: Melting method is used for high quality single crystals syntheses. For the $x = 0$ sample, the raw materials Mn, Bi, and Te were mixed in the molar ratio of 1:2:4 and carefully grounded in a glove box with argon protection. Then, the mixture was sealed inside a quartz ampule. The ampule was placed in a furnace and heated to 900°C in 12 h and maintained for 12 h. Then, the ampule was cooled to 600°C in 60 h and annealed over one week. The last step was to cool down to 590°C in 30 h and anneal for 3 days, followed by quenching in air. The high-quality single crystal flakes were easily exfoliated with a razor blade. For the other crystals with different values of $x$, Ge was used to replace Mn in the determined molar ratio. The growing process was exactly the same. Samples grown by this method were examined to be pure, without $Bi_2Te_3$ impurities.

XRD characterization: XRD measurements were performed on a BRUKER D8 ADVANCE diffractometer at room temperature. The cleaved (001) surfaces of $Mn_{(1-x)}Ge_xBi_2Te_4$ were carefully set to be parallel with the sample stage, and the beta-filter Ni with Ge004 beam path are applied to remove the Kα2 wavelength of Cu radiation.

HAADF-STEM characterization: For HAADF-STEM characterization, the Zeiss Crossbeam 550 Focused Ion Beam (FIB)-SEM TEM is used for preparing TEM lamellae, and the characterization was conducted on a probe and image corrected FEI Titan Themis Z microscope equipped with a hot-field emission gun working at 300 kV.

ARPES measurements: ARPES experiments are performed at Beihang University. The photoelectrons are excited by He-Iα light (21.2 eV) and the Scientia DA30L analyzer is used for measurements. The energy resolution is approximately 10 meV, and the momentum resolution is 0.01 Å$^{-1}$. The samples are cleaved and measured in an ultrahigh-vacuum chamber below $5^{-11}$ Torr.

Magnetic and transport measurements: The magnetic measurements were performed in a vibrating sample magnetometer (VSM) supplied in a physical property measurement system (Quantum Design). The measured samples were about 0.5 mm in thickness and 2 mm in side length. The electrical resistivity was measured using a physical property measurement system (Quantum Design), and the lowest temperature and highest magnetic field were 2 K and 9 T, respectively. The measured samples were about 0.5 mm in thickness and 2 mm in side length. Standard Hall-bar contacts and the four-probe method were used for in-plane resistivity $\rho_{xx}$ and Hall resistivity $\rho_{xy}$ measurements. Silver paste was used on the cleaved sample surface for immobilization and electron conduction.

First-principles calculations: The first-principles calculations are performed using the Vienna ab initio simulation package within the projector augmented wave method and the generalized gradient approximation of the Perdew-Burke-Ernzerhof exchange-correlation functional [27-29]. The plane-wave basis with an energy cutoff of 360 eV and 5×5×5 and 5×5×3 Γ-centered k-point meshes are adopted for FM and AFM/PM calculations, respectively. The doping ratio is set to $x$ = 0.44 in $Mn_{1-x}Ge_xBi_2Te_4$, which means 4 Mn atoms are substituted with Ge atom in a 3×3 supercell of $MnBi_2Te_4$ to simulate the actual situation of $x$ = 0.46. Employing the experimental lattice constants, the crystal structure is relaxed with van der Waals correction until the residual forces on each atom is less than 0.01 eV/Å. The SOC effect is considered in our calculations. The special quasi-random structure is adopted to simulate spin disorder PM states in a 3×3 supercell [30]. Such method offers the optimal level of randomness for a finite-sized supercell to ensure the closest match of correlation functions with those of the infinite alloy, suitable for both scenarios of composition disorder and spin disorder. A tight-binding Hamiltonian based on the maximally localized Wannier functions is constructed to further calculate the surface states and WCC using the WannierTools package [31,32].

## III. RESULTS

### 1. Crystal synthesis and intrinsic defect regulating

Aiming at regulating the Bi/Mn antisite defects in MBT, we adopted the strategy of element replacement. The MBT family has the same structure as its close relatives $GeBi_2Te_4$, $SnBi_2Te_4$ and $PtBi_2Te_4$. We selected Ge as the dopant due to the smallest lattice mismatch. The crystal structure of Ge doped MBT have a layered rhombohedral crystal structure composed of repeated septuple layers (SLs) of Te–Bi–Te–X–Te–Bi–Te (X = Mn/Ge) with the space group $R\bar{3}m$ (Fig. 1(a)) [20,33]. A series of MGBT single crystals with different doping level $x$ were synthesized using the melting method (see Methods). The ratio of (Mn+Ge):Bi:Te is nearly 1:2:4 for samples with different amounts of Ge substitution, as confirmed by the energy dispersive spectra (EDS) (see Supplemental Material Fig. S1 and Table S1 [34]). Consistent with its stoichiometric crystal structure, the substitutional doping mainly occurred between Mn and Ge atoms. Fig. 1(b) displays the X-ray diffraction (XRD) patterns for single crystals with diverse $x$ values. The position shift of the $(00\bar{2}7)$ diffraction peak indicates a slight lattice expansion of the $c$-axis as $x$ value increases, suggesting that there are no other structural changes. The upper panel of Fig. 1(c) shows optical images of the obtained MGBT single crystals used for further measurements, which all have flat and shiny surfaces. Low energy electron diffraction (LEED) tests were then performed before ARPES measurements, which confirmed the same (001) terminations of single crystals (lower panel of Fig. 1(c)).

Now we consider the Ge doping effect on regulating the antisite defects. With appropriate chemical potential (see Supplemental Material Fig. S3 [34]), we have calculated the formation energies of antisite defects in MBT ($x = 0$), GeBi$_2$Te$_4$ (GBT, $x = 1$), and the moderately doped sample MGBT (4/9 Ge in a 3×3 supercell, near $x = 0.46$), where only $Bi^+_{Mn/Ge}$ defects (Bi substitutes the Mn/Ge site) and $(Mn/Ge)^-_{Bi}$ (Mn/Ge substitutes the Bi site) defects are considered since they are the dominant defect types and also the critical factors related to the controversies mentioned above [17,18,24,26,35,36]. On the one hand, the formation energy and concentration of defects depend on the position of Fermi level. When the Fermi level is located in the gap, for both MBT (left panel in Fig. 1(d)) and GBT (right panel in Fig. 1(d)), $Bi^+_{Mn/Ge}$ defects are very stable even with negative formation energies, while $(Mn/Ge)^-_{Bi}$ defects have high positive formation energies. This is consistent with the heavy *n*-type doping behavior for both experimentally synthesized MBT and GBT single crystals. For MGBT (middle panel in Fig. 1(d)), the formation energies of $Bi^+_{Mn/Ge}$ [$(Mn/Ge)^-_{Bi}$] defects are increased to positive [decreased]. Such changes can be related to the changes of strength of bonding and position of valence band maximum (VBM). Roughly speaking, MBT have stronger bonding than GBT (see Supplemental Material Fig. S5 and S6 [34]), and the VBM of MGBT after alloying is higher than both MBT and GBT instead of intermediate between the two systems (see Supplemental Material Fig. S8 and Table S2 [34]). The higher VBM will increase the *n*-type defect formation energy but decrease the *p*-type defect formation energy. Consequently, the doped Ge could suppress the *n*-type doping effect caused by $Bi^+_{Mn/Ge}$ defects. On the other hand, the true Fermi level position is determined by the balance of *n*-type and *p*-type defects with the lowest formation energies. Near the conduction band minimum (CBM), for MBT and GBT, the still large difference in formation energy between *n*-type $Bi^+_{Mn/Ge}$ defect and *p*-type $(Mn/Ge)^-_{Bi}$ defect leads the Fermi level to be pinned deep into the conduction bands (CBs); while for MGBT, the small formation energy difference between *n*-type and *p*-type defects make the Fermi level be just pinned to the edge of the CBs (black dashed lines in Fig. 1(d)). Finally, let's discuss the changes of defects concentrations after doping. Around the real situation of Fermi level (*e.g.*, near the black dashed line in Fig. 1(d)), the $Bi^+_{Ge}/Ge^-_{Bi}$ defect formation energy in GBT is much lower than that of $Bi^+_{Mn}/Mn^-_{Bi}$ in MBT, which leads to higher defects concentrations in GBT. After doping, both $Bi^+_{Ge}$ and $Ge^-_{Bi}$ defects formation energies are increased with their concentrations reduced compared to the pure GBT; while compared to MBT, the $Bi_{Mn}^{+}$ formation energy is slightly decreased with its concentration increased but the $Mn_{Bi}^{-}$ concentration is decreased. The reduction of $Mn_{Bi}^{-}$ concentration may be beneficial for the observation of magnetic nontrivial gap. Based on all these calculated results, the Ge doping in MBT can effectively regulate the ratio of antisite defects and further suppress the *n*-type behavior.

Scanning transmission electron microscopy (STEM) was conducted on MBT, GBT, and $Mn_{0.54}Ge_{0.46}Bi_2Te_4$ single crystals to probe the defect regulations at the atomic scale. From the high-angle annular dark-field (HAADF) images (left panel in Fig. 1(e)), it can be clearly seen that there are a large number of $Bi_{Ge}^{+}$ antisite defects in GBT, which is reflected in the presence of numerous Bi atoms with high-contrast in the middle Ge atomic layer. In contrast, these high-contrast Bi atoms are essentially invisible in MGBT, which suggest that the $Bi_{Ge}^{+}$ was significantly reduced in MGBT sample. For better visualize, strain mappings from geometric phase analysis (GPA) are overlaid on the HAADF-STEM image (middle panel of Fig. 1(e)), in which defect-related local strains along the [001] direction are marked by different colors [37]. Since the MnTe middle layer in MBT is squeezed by the external $Bi_2Te_3$ layer, the replacement of Mn/Ge atoms by excessive Bi atoms will release compressive stress (red) in the middle MnTe layer, which is the case with $Bi_{Mn/Ge}^{+}$ antisite defects. Conversely, $(Mn/Ge)_{Bi}^{-}$ antisite defects will suppress the tensile stress (blue). It can be clearly seen that these strains are unevenly distributed in MBT and GBT with weakened compressive stress. However, in MGBT, the compressive stress is well preserved and the stress distribution is more uniform. Therefore, the positive $Bi_{Mn/Ge}^{+}$ antisite defects in MBT and GBT are found to commonly exist, but they are suppressed in $Mn_{0.54}Ge_{0.46}Bi_2Te_4$, which is consistent with the formation energy calculation results.

## 2. ARPES measurements on MGBT

To investigate the effect of Ge doping on the electronic band structure of MBT, ARPES measurements were carried out. As shown in Fig. 2(a) and 2(b), The bulk band gap of MBT is estimated to be about 200 meV, with its Fermi level located across the bulk CBs, which agrees well with previous reports [38]. It is surprisingly found that the Dirac point of $Mn_{0.54}Ge_{0.46}Bi_2Te_4$ sample is nearly located at its Fermi level (highlighted in Fig. 2(b) and its diagram in Fig. 2(a)). In the case of the other four samples ($x$ = 0, 0.26, 0.67, 1), they keep retaining the *n*-type doping band structure, among which, the *n*-type behavior is reduced for $x$ = 0.26 and $x$ = 0.67 with their Dirac points closer to the Fermi level. Together with the above-mentioned results of antisite defects (Fig. 1(d) and (e)), it is reasonable to believe that the effective regulation of antisite defects ensures the intrinsic topological properties of $Mn_{0.54}Ge_{0.46}Bi_2Te_4$ observed near the Fermi level, as desired for MBT.

A detailed analysis then was made for Mn$_{0.54}$Ge$_{0.46}$Bi$_2$Te$_4$ (Fig. 2(c) to (j)). The spectra along the $k_x$ and $k_y$ directions show a clear Dirac-cone-like feature with linear dispersion (Fig. 2(c) and (d)). There is still a finite spectral weight at Γ point near the Fermi level, which is shown in the corresponding constant-energy contours (Fig. 2(e)). To identify the electronic states at the Fermi level, two momentum distribution curves (MDCs) were selected from the raw data (Fig. 2(g)). Based on the peak fitting results of the MDC1 curve ($E$-$E_F$=-20 meV), two strong peaks (red) and one weak peak (blue) are attributed to the topological surface states (TSSs) and VBM, respectively (Fig. 2(h) and 2(i)). Compared with MDC1, the intensity of the VBM obviously increases in the MDC2 curve ($E$-$E_F$=-60 meV), which is located a little away from the Fermi level. Fig. 2(j) provided more plots of the intensity of the VBM, in which the VBM suddenly decays as it approaches the Fermi level. This confirms that the Fermi level of Mn$_{0.54}$Ge$_{0.46}$Bi$_2$Te$_4$ is located in the bulk band gap and the states here are mainly contributed by the TSSs.

## 3. Magnetic and transport measurements of MGBT

Magnetic and transport measurements on MGBT were performed to study its intrinsic magnetic topological phenomena and TPTs behaviors. Fig. 3(a) shows the field-cooled (FC) and zero-field-cooled (ZFC) curves of samples from $x$ = 0 to 0.72, in which all samples show antiferromagnetic (AFM) ground state [11,21]. The Néel temperature ($T_N$) decreases from 25.4 K to 7.4 K as the Ge doping level increases from $x$ = 0 to 0.72. When an out-of-plane magnetic field is applied, all the samples undergo two transitions as the magnetic field increases: from AFM to canted-antiferromagnetic (CAFM) and from CAFM to ferromagnetic (FM), with the transition magnetic fields marked as $B_{sf}$ (spin flop field) and $B_{sa}$ (saturation field), respectively (Fig. 3(b)). By plotting $T_N$, $B_{sf}$, and $B_{sa}$ vs. the Ge doping level, the weakening of AFM order in MGBT as the Ge contents increase can be confirmed (Fig. 3(c)). The Hall transport measurements on MGBT shows that the Hall resistivity slope of Mn$_{0.54}$Ge$_{0.46}$Bi$_2$Te$_4$ is much larger than those of other samples (Fig. 3(d)). The carrier concentration $n$ and mobility $\mu_h$ can be derived from the Hall measurements. Again, Mn$_{0.54}$Ge$_{0.46}$Bi$_2$Te$_4$ has the highest mobility and the minimum carrier concentration (the inset of Fig. 3(d)). Combining with the ARPES measurements, these results can be explained by the successful tuning of the Fermi level towards the bulk gap and the surface Dirac point in Mn$_{0.54}$Ge$_{0.46}$Bi$_2$Te$_4$.

Fig. 3(e) presents the temperature dependence of the in-plane resistivity ($\rho_{xx}$), in which the samples of $x$ = 0, 0.26, 0.67 show metallic behaviors with a remarkable kink at their $T_N$. This is a normal phenomenon in AFM materials which can be related to the spin-fluctuation induced spin scatterings [39]. In contrast, a metal-insulator transition is observed for Mn$_{0.54}$Ge$_{0.46}$Bi$_2$Te$_4$ at its $T_N$. This transition can be explained by a possible TPT from a STI to an MTI at $T_N$: above $T_N$, the paramagnetic (PM) $Mn_{0.54}Ge_{0.46}Bi_2Te_4$ may exhibit a STI with gapless Dirac surface states, which contributes to metallic behavior; below $T_N$, the established AFM order turns the system into an MTI and opens a magnetic gap, leading to insulating behavior [16]. Besides the contribution of surface conductivity, it is noted that the bulk conductivity may also contribute the metallic behavior of $Mn_{0.54}Ge_{0.46}Bi_2Te_4$ at high temperature. First, the bulk band gap may decrease at the PM order. Second, the Fermi surface will be broadened at high temperature. In addition, antisite defect regulation in different sample areas may be uneven, causing the Fermi level to contact the bottom of the CBs or the top of the VBs.

Other possible mechanisms for the metal-insulator transition accompanying by a magnetic phase transition can be excluded, such as the orbital-selected Mott mechanism and the Slater mechanism [40-43]. Here we did not observe any Hubbard-related flat bands or band folding effect from the ARPES results, which are essential features for these two mechanisms, respectively. Combining the data from magnetization and Hall resistivity ($M$-$T$ and $\rho_{xy}$-$T$), it is believed that the TPT from a STI to an MTI in $Mn_{0.54}Ge_{0.46}Bi_2Te_4$ is induced by the temperature-dependent magnetic transition from a PM to an AFM state. The metal-insulator transition is not observed in other Ge doped samples due to their heavy $n$-type doping fact with their transport properties dominated by the CBs.

In-plane resistivity vs. temperature ($\rho_{xx}$-$T$) measurements under different out-of-plane magnetic fields were also carried out on $Mn_{0.54}Ge_{0.46}Bi_2Te_4$ (Fig. 3(f)). The metal-insulator transition behavior is almost unchanged between 0-2 T with its antiferromagnetic ground state maintained. Between 3 T and 4.5 T where $Mn_{0.54}Ge_{0.46}Bi_2Te_4$ undergoes a magnetic transition from AFM to CAFM, the metal-insulator transition temperature decreases from around 15 K to 9 K. This phenomenon is caused by the reduced effective out-of-plane magnetism in the CAFM state, and meanwhile the magnetic gap near the Fermi level is reduced. At 6 T when $Mn_{0.54}Ge_{0.46}Bi_2Te_4$ becomes ferromagnetic, the metal-insulator transition disappears. The recovery of metallic behavior at higher magnetic fields is consistent with the gapless type-II WSM phase predicted for the MBT system when its AFM order is coerced into FM order [16]. One of the experimental characteristics of type-II WSM is its unsaturated magnetoresistance effect [44]. To support this, magnetoresistance measurement on the $Mn_{0.54}Ge_{0.46}Bi_2Te_4$ sample shows that there is a nearly-linear unsaturated magnetoresistance when the sample is completely transformed into the FM state up to the largest applied magnet field of 9 T, indicating a WSM phase (see Supplemental Material Fig. S11 [34]). As a result, another TPT from MTI to WSM is revealed on the $Mn_{0.54}Ge_{0.46}Bi_2Te_4$ sample, which is accompanied by the AFM to FM transition.

## 4. DFT calculated results for MGBT

To verify the above two magnetism-induced TPTs in $Mn_{0.54}Ge_{0.46}Bi_2Te_4$, density-functional theory (DFT) calculations on the doping level of $x = 0.44$ (4/9 Ge in a 3×3 supercell) were performed. The bulk and surface bands of $Mn_{0.54}Ge_{0.46}Bi_2Te_4$ with spin-orbit coupling (SOC) and without SOC in different magnetic states are demonstrated (see Supplemental Material Fig. S4 and S12 [34]). Similar to the pure MBT, $Mn_{0.54}Ge_{0.46}Bi_2Te_4$ also has an AFM ground state. Compared to the large band gap of around 200 meV in AFM MBT, the band-gap size of AFM $Mn_{0.54}Ge_{0.46}Bi_2Te_4$ has a significant reduction to 37 meV, which is mainly caused by the bulk band evolution with non-magnetic Ge doping (Fig. 4(a)). The topological index of AFM $Mn_{0.54}Ge_{0.46}Bi_2Te_4$ is calculated as $Z_2 = (1,000)$, same as MBT, revealing that the AFM TI nature is well maintained (see Supplemental Material Fig. S13 [34]). Fig. 4(b) presents the topological surface states of the (001) and (100) surfaces in which a magnetic gap of about 10 meV opens at the Dirac point on the (001) surface but the gapless Dirac states are retained on the (100) surface. The band structure of $Mn_{0.54}Ge_{0.46}Bi_2Te_4$ in the PM state is shown in Fig. 4(d) and (e), similar to the AFM ground state, the bulk band under PM state still keeps the insulating feature. But with disordered and net zero magnetic moments, both (001) and (100) surfaces have gapless Dirac surface states. Therefore, the PM $Mn_{0.54}Ge_{0.46}Bi_2Te_4$ is a STI. For the FM state of $Mn_{0.54}Ge_{0.46}Bi_2Te_4$ where all the magnetic moments align to the $z$ direction, the bulk gap is closed with a pair of band crossing occurring along the Z–Γ–Z high-symmetry line (Fig. 4(c) and (g)). Wannier charge center evolution reveals that the Chern number is $C = 1$ and $C = 0$ on the $k_z = 0$ and $k_z = \pi$ planes respectively (see Supplemental Material Fig. S13 [34]), which confirms that the FM $Mn_{0.54}Ge_{0.46}Bi_2Te_4$ is a WSM. The Weyl points can be reflected by the surface states calculations (Fig. 4(h)), and the Fermi arc connecting two Weyl points in the momentum space is clearly seen. (Fig. 4(i) and S14, Supporting Information [34]).

These calculated electronic structures can match well with our transport measurements demonstrating the TPTs. The sample of $Mn_{0.54}Ge_{0.46}Bi_2Te_4$ has the optimal doping ratio, which largely suppresses the $n$-type defects and positions the Fermi energy precisely within the bulk bandgap, thereby minimizing the contribution of bulk conductivity to the transport properties. Fig. 4(f) shows the entire phase transition process. Above the Néel temperature, $Mn_{0.54}Ge_{0.46}Bi_2Te_4$ with the PM state is a STI, whose gapless topological surface states contribute to the metallic transport. With decreasing the temperature to a magnetic transition from PM to AFM (ground state), a gap is opened on the (001) surface due to the accompanying TPT from STI to WTI, which results in the insulating feature. When a high magnetic field along $z$ is applied, turning the system into a FM state, the emerging WSM phase restores the metallic behavior in transport, as also revealed by unsaturated quasi-linear magnetoresistance effect. It is worth noting that the two Weyl points are located very close to the Γ point and are also very close to each other, which results in a relatively small Berry curvature.

Therefore, it is difficult to directly observe the negative magnetoresistance caused by the abnormal chiral transport in the experiments.

## IV. CONCLUSION

In this work, we propose a promising strategy to effectively regulate the antisite defects and suppress the heavy *n*-type doping in MBT, which is critical for experimental realizations of intrinsic magnetic topological phenomena and precise magnetic control of TPTs. By substitutional doping Ge in MBT, the ratio of intrinsic antisite defects can be efficiently regulated in MGBT, and the Fermi level can be tuned to the bulk gap and close to the surface Dirac point in moderately doped $Mn_{0.54}Ge_{0.46}Bi_2Te_4$ sample, as confirmed by our defect formation energy calculations, STEM and ARPES measurements. Especially, the $Mn_{Bi}^-$ concentration can be effectively reduced after Ge doping according to our calculations, which is important for the realization of magnetic nontrivial gap. Remarkably, two magnetism-dependent TPTs from STI to MTI and then to WSM are observed in our transport measurements and electronic structure calculations, as accompanied by the metal-insulator-metal transitions together with a PM-AFM-FM magnetic phase transition process. Our work paves the way for achieving intrinsic and controllable magnetic topological quantum states and controllable and continuous TPTs towards future spintronic applications. For instance, the preparation of thin films from the discovered optimal $Mn_{0.54}Ge_{0.46}Bi_2Te_4$ material may serve as an ideal platform to access high-temperature QAHE and AI state, and its heterostructure with superconductors promises the realization of Majorana fermion for the potential quantum computing.

## Acknowledgements

The authors are grateful to the Analysis & Testing Center of Beihang University for the facilities, and the scientific and technical assistance. The authors thank S.-H. Wei and X. Yan for helpful discussions. This work was supported by the National Natural Science Foundation of China (Grant Nos. 52473287, 12474217, 12004030, 12004021 and 12274016), Beijing Natural Science Foundation (Grant No. 1242023), the Fundamental Research Funds for the Central Universities (Grants YWF-23SD00-001 and YWF-22-K-101) and the National Key R&D Program of China (2018YFE0202700 and 2022YFB3403400).

## Author Contributions

Y.D. and W.H. planned the experimental project. S.X., M.Y., and J.Z. conducted ARPES experiments and analysed the data. S.X. and H.X. made and characterized single crystals. N.C. performed the HAADF-STEM experiments. Z.S. and J.W. conducted the DFT calculations. S.X., Z.S., H.F., J.W, W.H. and Y.D. wrote the paper. Y.D. supervised this work. All authors discussed the results and commented on the manuscript.## Data Availability Statement

The data that support the findings of this study are available from the corresponding author upon reasonable request.

# Figures

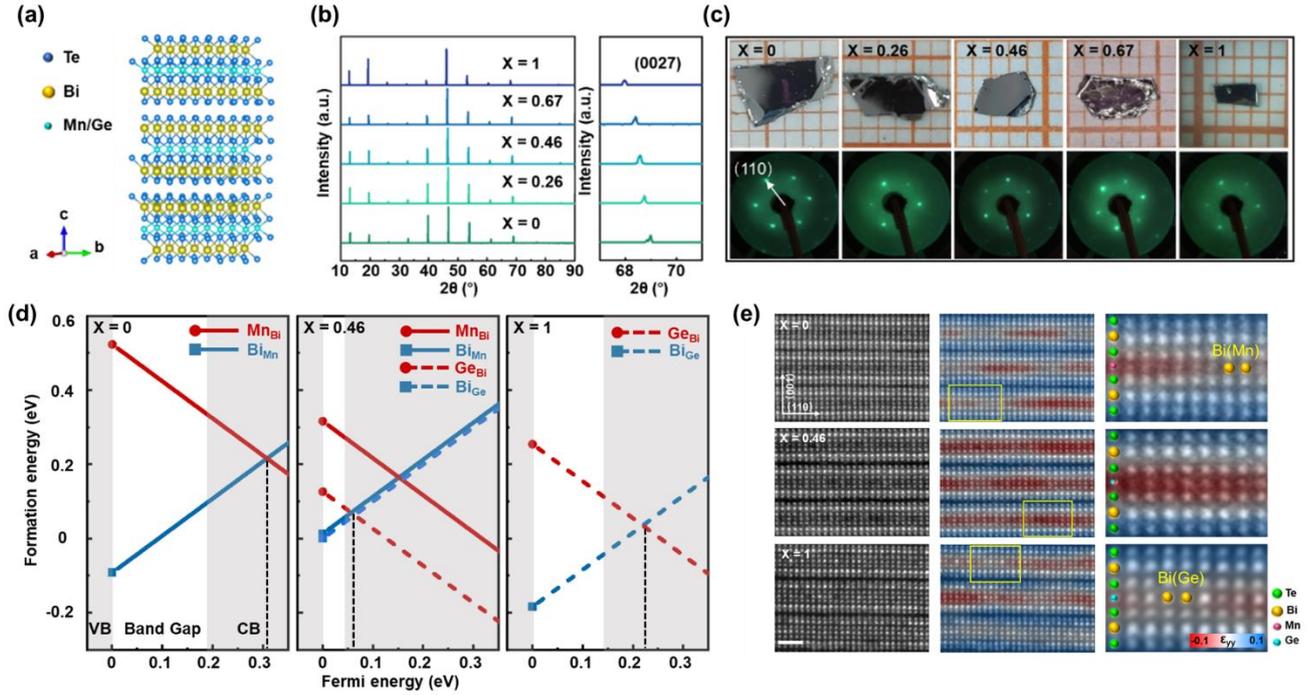

**FIG 1. Sample characterizations and identification of antisite defects.** (**a**) Schematic diagram of the crystal structure of MGBT. (**b**) XRD patterns for samples with different Ge doping level. (**c**) Optical images and LEED images of different samples, with the white arrow indicating the [110] in-plane direction. (**d**) Calculated results of the defect formation energy of MBT ($x = 0$), MGBT (near $x = 0.46$) and GBT ($x = 1$). The red (dashed) line represents the $Mn_{Bi}$ ($Ge_{Bi}$) antisite defect, while the blue (dashed) line represents the $Bi_{Mn}$ ($Bi_{Ge}$) antisite defect. The black dashed line represents the pinned, possibly true Fermi level position. (**e**), HAADF images of MBT ($x = 0$), $Mn_{0.54}Ge_{0.46}Bi_2Te_4$ ($x = 0.46$), and GBT ($x = 1$). Scale bar is 1 nm. The middle panel shows the GPA results, in which the red (blue) color marks the contraction (expansion) region. The right panel shows the zoomed areas in the yellow box of the middle panel. The larger areas of STEM image and EDS mapping results are shown in Fig. S2.

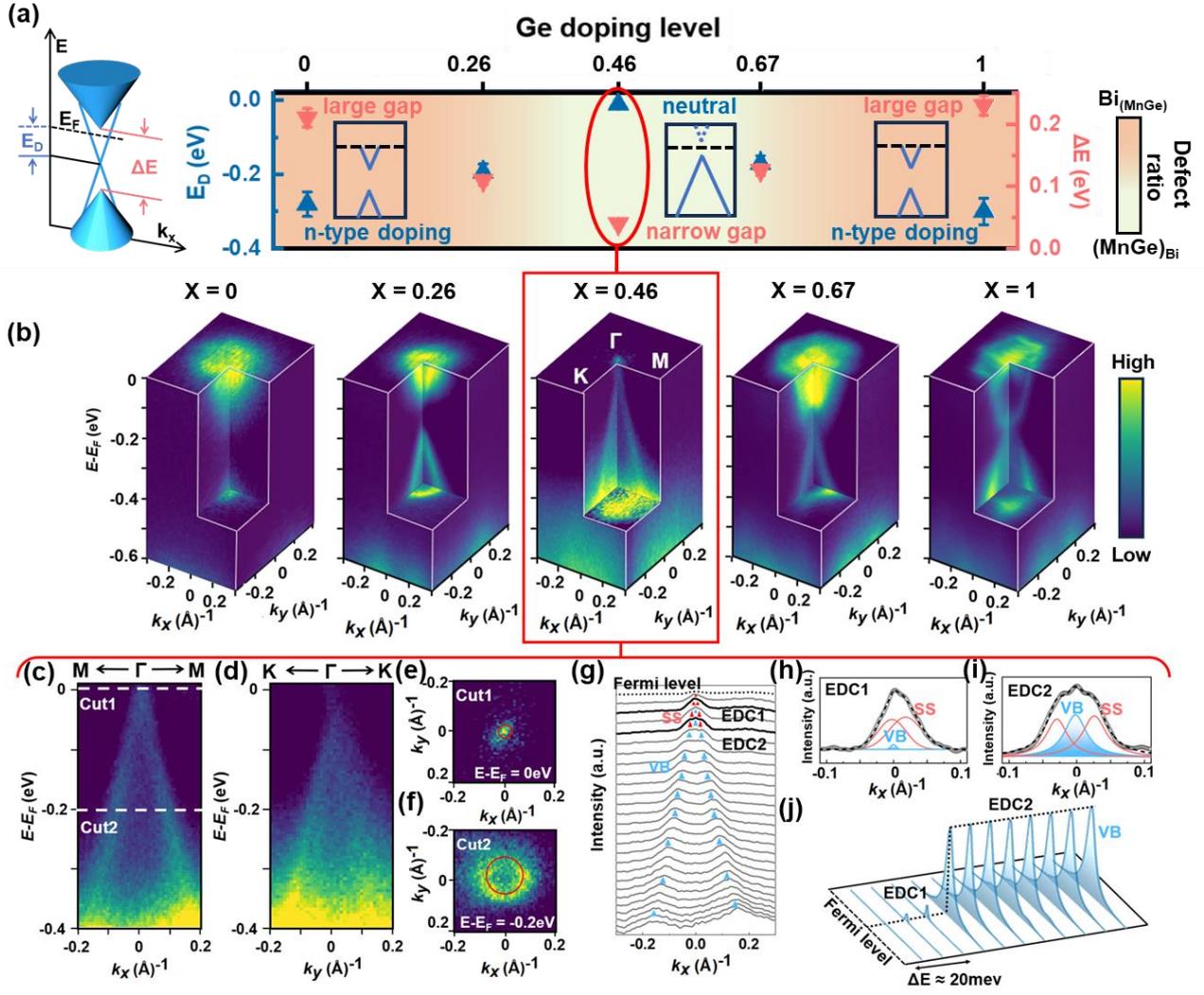

**FIG 2. ARPES measurements of MGBT samples.** (**a**) Schematic diagram of the energy band of MGBT, where $E_D$ means the energy of Dirac point relative to the Fermi level ($E_F$), and $\Delta E$ means the value of the bulk band gap. (**b**) ARPES measurements of different Ge doped samples, with the measurement temperature set to 7.7 K and the beam energy at 21.2 eV. (**c**) and (**d**) Detailed ARPES data on $Mn_{0.54}Ge_{0.46}Bi_2Te_4$ along the M–Γ–M and K–Γ–K paths. (**e**), (**f**), constant-energy contours at the Fermi level (cut1) and -0.2 eV (cut2). (**g**) Momentum distribution curves derived from the ARPES spectra in (**c**), with the blue arrows marking the valence band (VB) and the red arrows marking the surface states (SS). (**h**) and (**i**) Lorentz peak fitting results of MDC1 (**h**) and MDC2 (**i**) picked from (**g**). (**j**) More plotted results of the intensity of the VB.

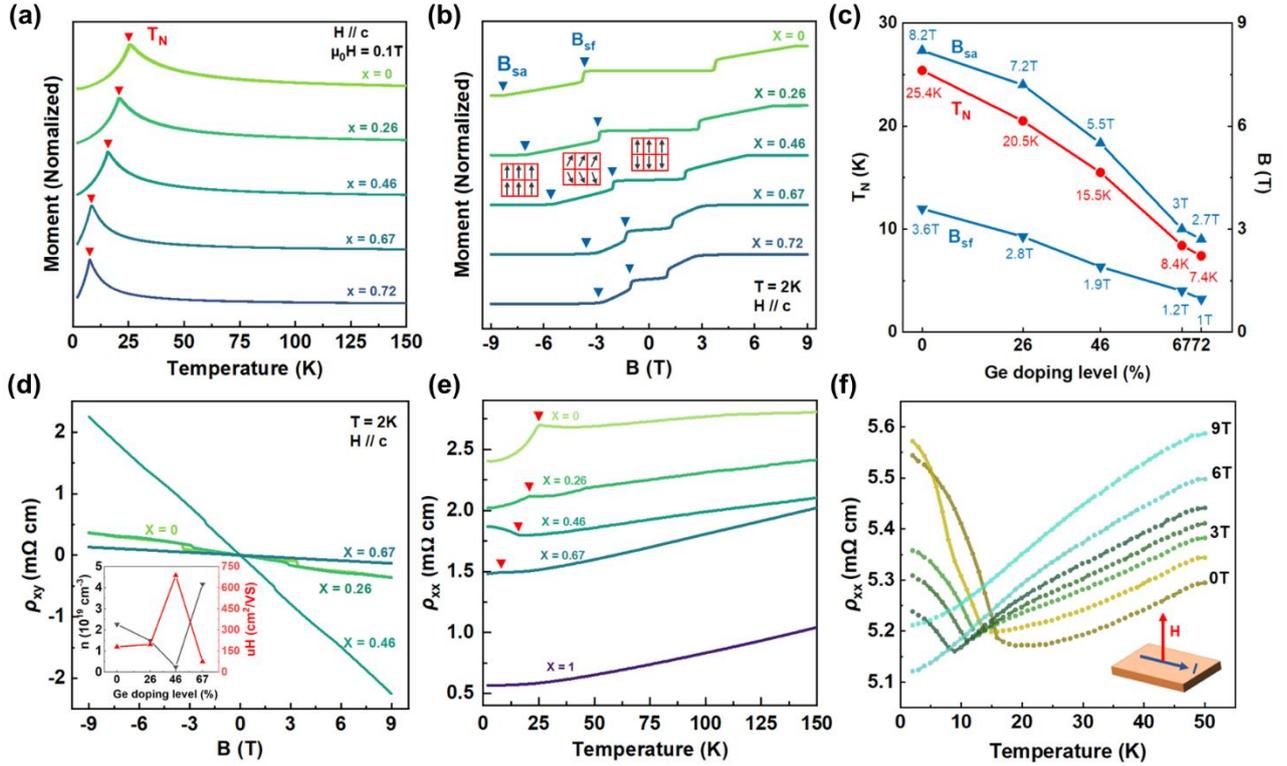

**FIG 3. Magnetic and transport measurements of Mn$_{1-x}$Ge$_x$Bi$_2$Te$_4$ samples.** (**a**) Temperature dependence of the magnetization (*M-T* curves) of the samples with different values of *x*, where the moment is normalized for better comparison. (**b**) *M-H* curves of samples (*T* = 2 K, *H* is the magnetic field strength and *H*//*c*). The insets schematically plot the AFM, CAFM, and FM states. The blue arrows mark the critical magnetic fields between these states. The magnetism data are all normalized, considering that the magnetic moment per unit decreases as the doping level increases, as illustrated in Fig. S10. (**c**) The relationships between $T_N$, $B_{sa}$, $B_{sf}$ and the Ge doping level. (**d**) Hall measurements on Mn$_{1-x}$Ge$_x$Bi$_2$Te$_4$ samples (*T* = 2 K, *H*//*c*), with the inset showing the carrier concentration *n* and mobility $\mu_h$ at different doping levels. (**e**) *ρ-T* curves of Mn$_{1-x}$Ge$_x$Bi$_2$Te$_4$ samples. (**f**) *ρ-T* curves of the *x* = 0.46 sample under different out-of-plane magnetic fields (0 T, 1 T, 2 T, 3 T, 3.7 T, 4.5 T, 6 T, 9 T).

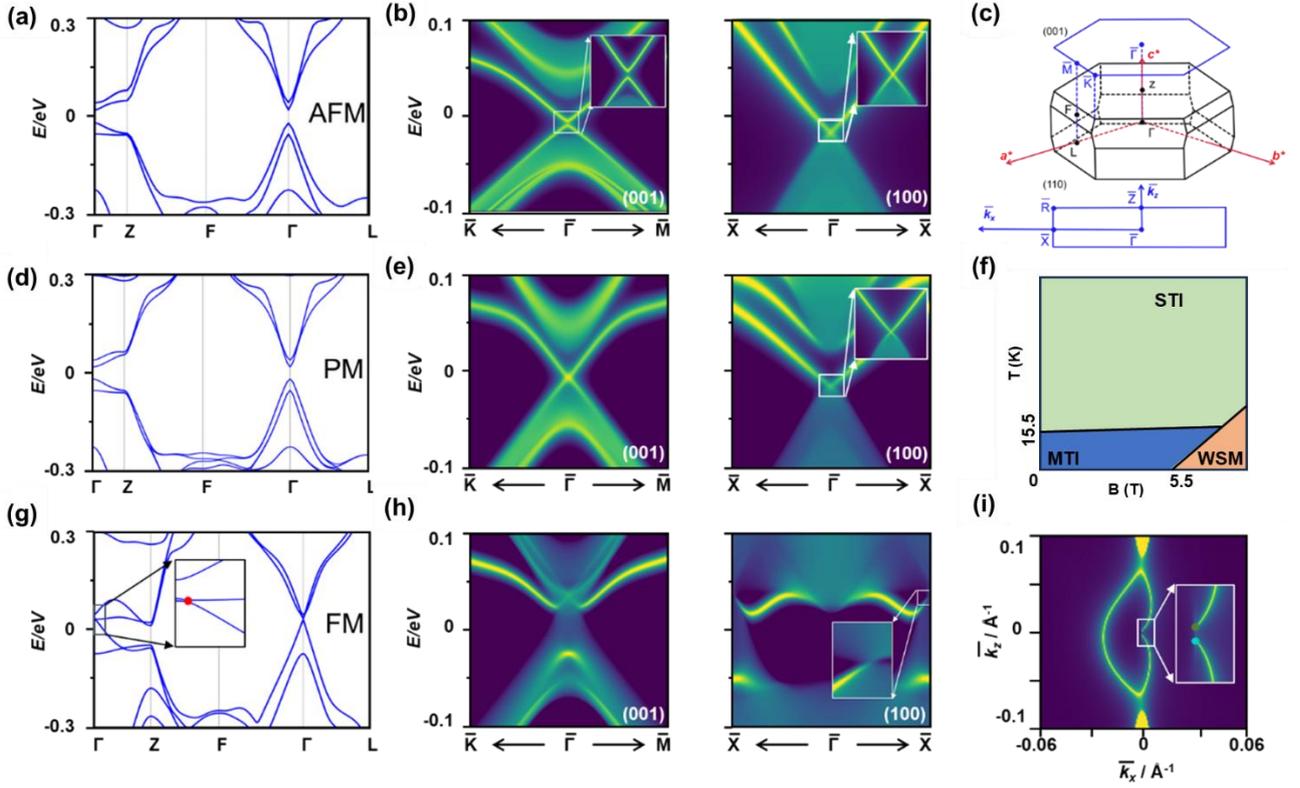

**FIG 4. DFT calculated results on Mn$_{0.54}$Ge$_{0.46}$Bi$_2$Te$_4$.** (**a**) (**d**) (**g**) Bulk band structures of Mn$_{0.54}$Ge$_{0.46}$Bi$_2$Te$_4$ with different magnetic states: AFM (**a**), PM (**d**), and FM (**g**). (**b**) (**e**) (**h**) Surface band structures of the (001) and (100) surfaces of Mn$_{0.54}$Ge$_{0.46}$Bi$_2$Te$_4$ with different magnetic states: AFM (**b**), PM (**e**), and FM (**h**). (**c**) Brillouin zone of Mn$_{0.54}$Ge$_{0.46}$Bi$_2$Te$_4$. (**f**) Phase diagram of Mn$_{0.54}$Ge$_{0.46}$Bi$_2$Te$_4$ with different magnetic field and temperature. (**i**) Fermi arc of Mn$_{0.54}$Ge$_{0.46}$Bi$_2$Te$_4$ in the FM state.

Supplementary Materials for

# Controllable and Continuous Quantum Phase Transitions in Intrinsic Magnetic Topological Insulator

Shengjie Xu[1,†], Zhijian Shi[1,†], Ming Yang[2], Jingwei Zhang[1], Hang Xu[1], Haifeng Feng[1], Ningyan Cheng[3], Jianfeng Wang[1,*], Weichang Hao[1,2,*], Yi Du[1,2,*]

[1] *School of Physics, Beihang University, Haidian District, Beijing 100191, China*
[2] *Analysis & Testing Center of Beihang University. Beihang University, Beijing 100191, China*
[3] *Information Materials and Intelligent Sensing Laboratory of Anhui Province, Key Laboratory of Structure and Functional Regulation of Hybrid Materials of Ministry of Education, Institutes of Physical Science and Information Technology, Anhui University, Hefei, Anhui, China*

*†Shengjie Xu and Zhijian Shi contributed equally to this work.*
*Correspondence authors. E-mail: yi_du@buaa.edu.cn; whao@buaa.edu.cn; wangjf06@buaa.edu.cn

# Supplementary Notes:

## Note 1. EDS / STEM results for the MGBT system.

The detailed EDS data for MGBT ($x = 0.46$) is shown in Figure S1, and a summary of the EDS mapping results for $Mn_{1-x}Ge_xBi_2Te_4$ is shown in Table S1. Consistent with its stoichiometric crystal structure, the doping mainly occurred between Mn and Ge atoms. It is also worth noting that the ratio of (Mn+Ge):Bi:Te in the $x = 0.46$ sample is mostly nearly 1:2:4 compared with the other samples, suggesting that the antisite defects were successfully regulated.

Fig. 1e in the main text was taken from a smaller area of Figures S2a to 2c, which shows the strain mappings from geometric phase analysis (GPA) overlaid on the large scale HAADF-STEM image of MBT ($x = 0$), MGBT ($x = 0.46$) and GBT ($x = 1$). Figure S2d shows the atomically resolved EDS mapping image of the MGBT ($x = 0.46$) sample. The distribution of elements can be clearly seen.

## Note 2. Detailed formation energy calculation for the MGBT system.

The formation energy calculation follows

$$\Delta H_{D,q}(E_F, \mu) = [E_{D,q} - E_H] + q(E_V + \Delta E_F) + \sum n_\alpha(\mu_\alpha^0 + \Delta\mu_\alpha), \quad (S1)$$

where $E_D$ and $E_H$ are the total energies of the host+defect and host-only supercells respectively, $E_V$ and $\Delta E_F$ are the energy of valence band maximum (VBM) and Fermi energy with respect to VBM respectively, $\mu_\alpha = \mu_\alpha^0 + \Delta\mu_\alpha$ is the chemical potential of the atom $\alpha$, $q$ and $n_\alpha$ are the charged defect valence and the defect atom numbers respectively. The potential alignment and image charge corrections are considered in the calculation process. Figure S3a shows the 3×3×2 supercell of MGBT as adopted in the defect formation energy calculations. The Mn and Ge atoms are randomly and uniformly distributed in the supercell structure. Figure S3b shows the feasible chemical potential range for formation of MBT as calculated under thermal equilibrium, which is represented approximately by a line segment between points A and B. The small stable region suggests the synthesis of these compounds can be challenging. However, according to the chemical potential calculation results of GBT in Figure S3c, we cannot find a range that allows the GBT to form stably, which means that GBT is a metastable structure and the growth conditions are more stringent than MBT. In our calculations of defects formation energies as functions of Fermi energy in the main text, the chemical potentials at the B point of MBT and at the A point of GBT are adopted. For the potential alignment, the work function calculations with slab models are adopted to compare the energy of VBM for MBT and GBT (Figure S4); while for the Ge doped structure, a method with core level alignment is employed.

Based on Equation (S1), the charged defect formation energy is related to two factors: one is the defect formation energy regardless of charge, and the other is the VBM. The first factor depends on the strength of bonding between Mn/Ge, Bi and Te atoms. Our calculations show that MBT has stronger bonding than GBT, as the $d$ orbitals of Mn overlap more with the $p$ orbitals of Bi and Te atoms than the $s$ orbitals of Ge (Figures S5 and S6). Therefore, doping Ge into MBT weakens the bonding of MBT. As a result, the $Bi_{Mn}$ ($Bi_{Ge}$) defects become more localized (delocalized) with a lower (higher) formation energy in MGBT than that in MBT (GBT); the similar results can apply to $Mn_{Bi}$ ($Ge_{Bi}$) defects (Figure S7). Regarding the second factor, our calculations show that the VBM of MGBT after alloying is higher than that of MBT and GBT (Table S2), and is not intermediate between the two systems. This is because the VBM after alloying is not evenly distributed in the Bi and Te layers, but rather resembles phase separation (Figure S8). The higher VBM will increase the $n$-type defect formation energy but decrease the $p$-type defect formation energy. The combined two factors mentioned above lead to the final results in Fig. 1d of the main text.

Another thing worth noting is that in our calculations as shown in Figure 1d of the main text, the positions of antisite defects in the 3×3×2 supercell are adopted in only one septuple layer which may slightly break the AFM order of MBT and MGBT. To check its influence, here two antisite defects are also adopted in both septuple layers preserving the AFM orders. Figure S9 shows the calculated formation energies under this situation, which revealing that except for the numerical value doubling, the main results remain basically unchanged.

**Note 3. Magnetism data without normalized.**

It needs to be mentioned that for better comparison, the magnetism data are all normalized in Fig. 3a and 3b of the main text. In fact, as the doping content increases, the magnetic moment per unit mass decreases, as shown in Figure S10.

**Note 4. Magnetoresistance results for the MGBT system.**

Figure S11 shows the magnetoresistance (MR) versus magnetic field of samples with different doping levels. Two types of transition features (marked by red arrows) can be identified in all four samples. They can be explained by two transitions from an AFM to CAFM and from a CAFM to FM caused by the external magnetic field respectively. Figure S11c shows the near-linear magnetoresistance of the sample with $x$ = 0.46 until 9T, which may be consistent with the Weyl semimetal properties. However, whether it is still unsaturated at higher magnetic fields requires additional data support.

**Note 5. Additional topological properties calculations.**

Figure S12 shows the energy band of $Mn_{0.56}Ge_{0.44}Bi_2Te_4$ in AFM and FM without spin-orbit coupling (SOC). The band inversion does not exist in both states, highlighting the importance of SOC in the formation of topologically non-trivial electronic structures.

Figure S13 shows the evolution of Wannier charge center (WCC) with AFM, PM, and FM states. The calculated topological invariants are $C = 1$ on $k_z = 0$ plane and $C = 0$ on $k_z = \pi$ plane for the FM state, $Z_2 = 1$ on $k_z = 0$ plane and $Z_2 = 0$ on $k_z = \pi$ plane for the AFM and PM states. It is worth noting that since the topological protection of $S$-symmetry only exist on the $k_z = 0$ plane, the discussion of the $Z_2$ invariant on $k_z = \pi$ plane is nonsense.

Figure S14 shows the evolution of surface Fermi arc for FM Weyl $Mn_{0.56}Ge_{0.44}Bi_2Te_4$ at different Fermi energies from 0.03eV to 0.05 eV to provide supporting information in Figure 4i of the main text.

**Supplementary Figures:**

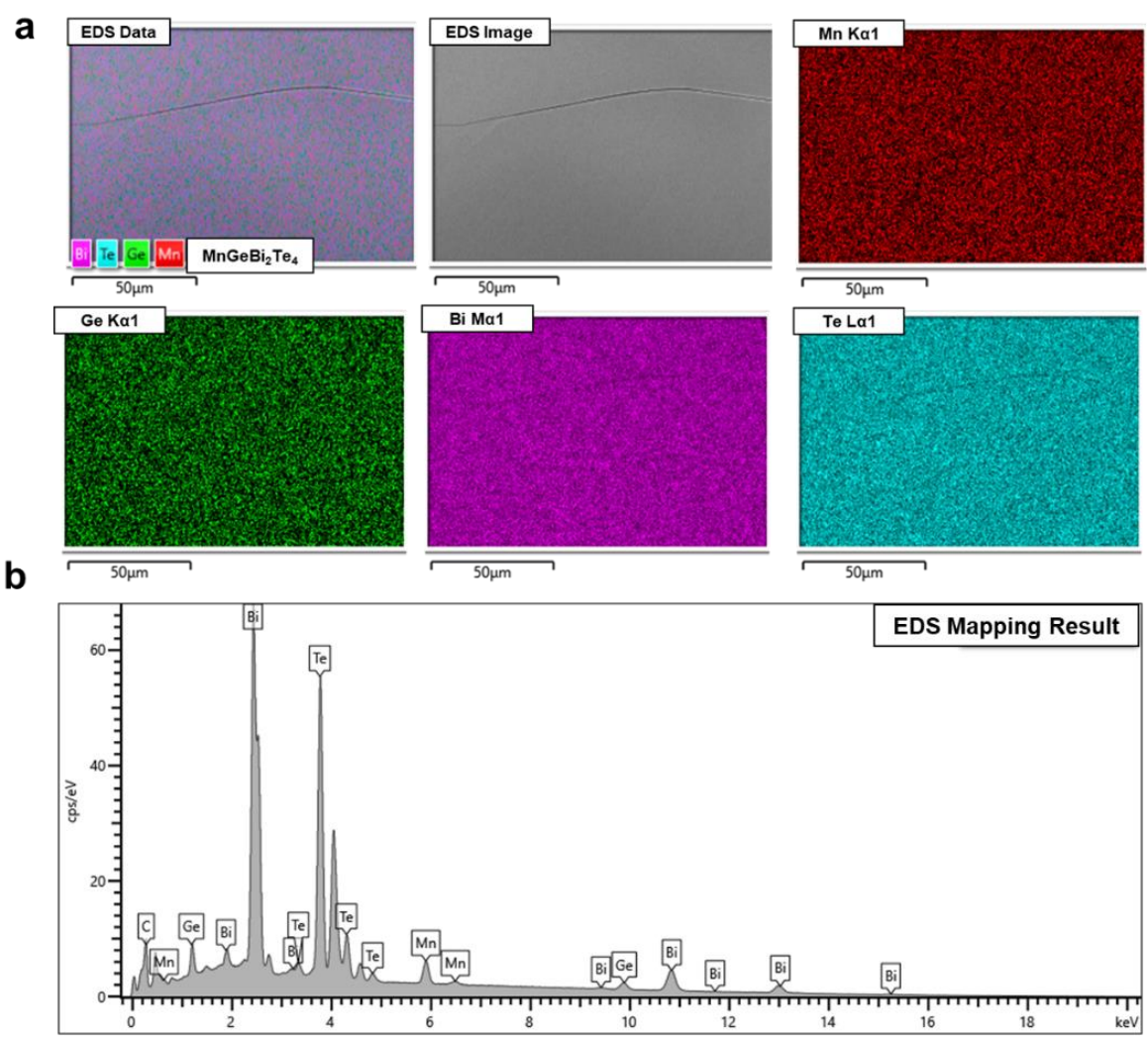

**Figure S1.** The detailed EDS data for MGBT ($x$ = 0.46). (a) Mapping images of MGBT ($x$ = 0.46). (b) The mapping spectrum of MGBT ($x$ = 0.46).

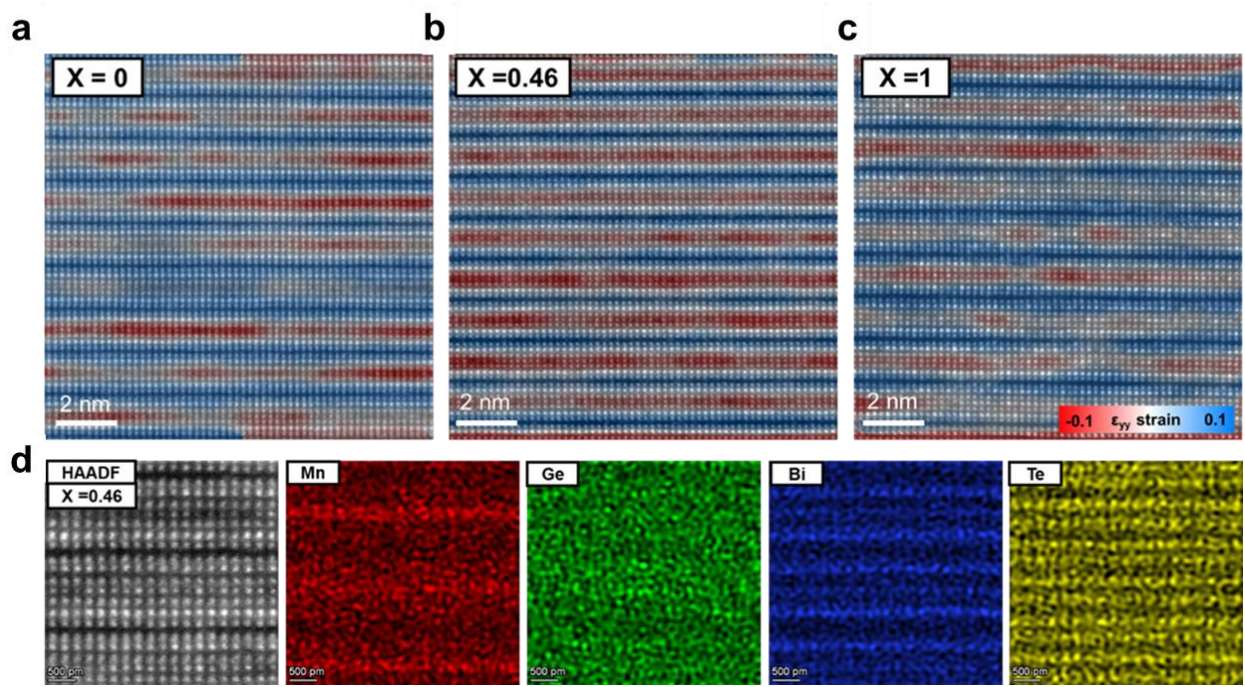

**Figure S2.** Strain mappings from geometric phase analysis (GPA) overlaid on the large-scale HAADF-STEM images of (a) MBT ($x = 0$), (b) MGBT ($x = 0.46$), and (c) GBT (x = 1). (d) Atomically resolved EDS mapping images of MGBT sample (x = 0.46).

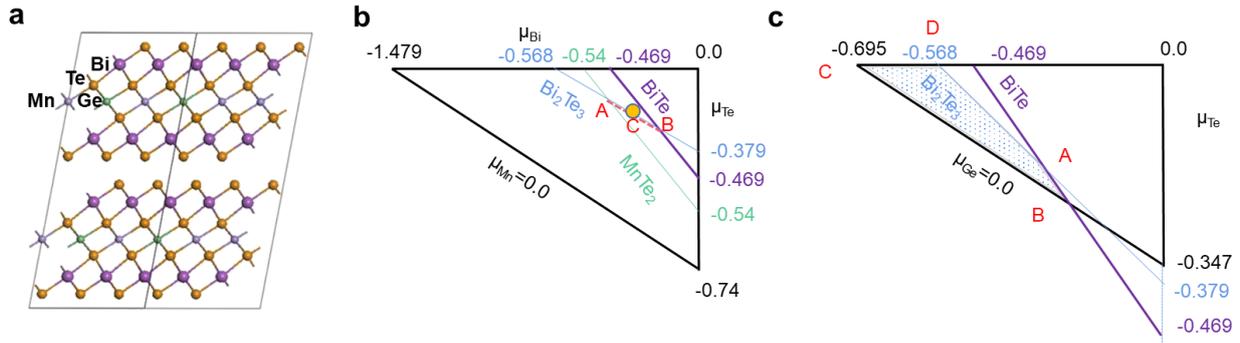

**Figure S3.** (a) Supercell structure of $Mn_{1-x}Ge_xBi_2Te_4$ in our defect formation energy calculations. Calculated chemical potential ranges for formation of (b) MBT and (c) GBT.

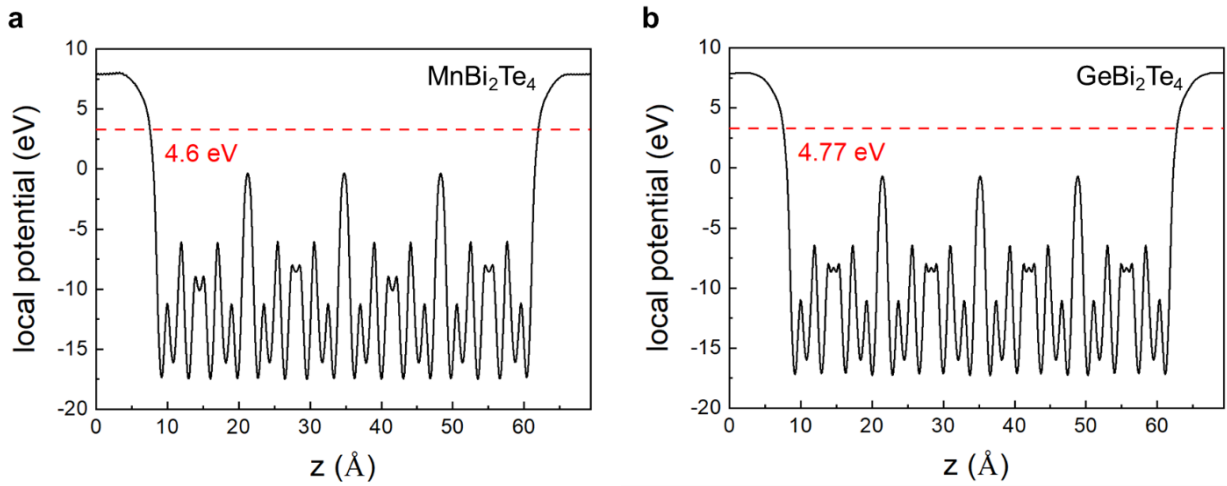

**Figure S4.** Calculated work functions from the slab model for (a) $MnBi_2Te_4$ and (b) $GeBi_2Te_4$, where the red dashed lines denote the Fermi energies.

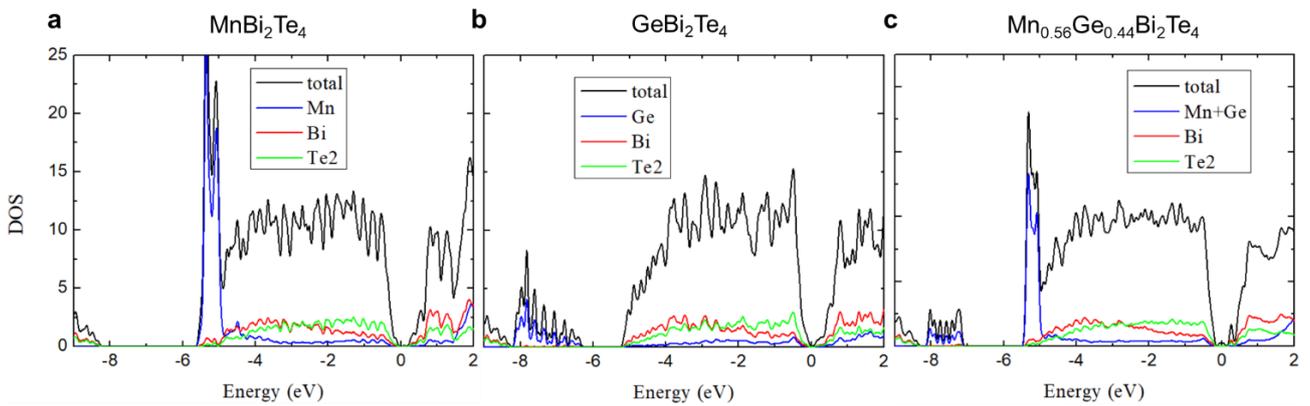

**Figure S5.** Density of states projected on atoms for (a) $MnBi_2Te_4$, (b) $GeBi_2Te_4$ and (b) $Mn_{0.56}Ge_{0.44}Bi_2Te_4$.

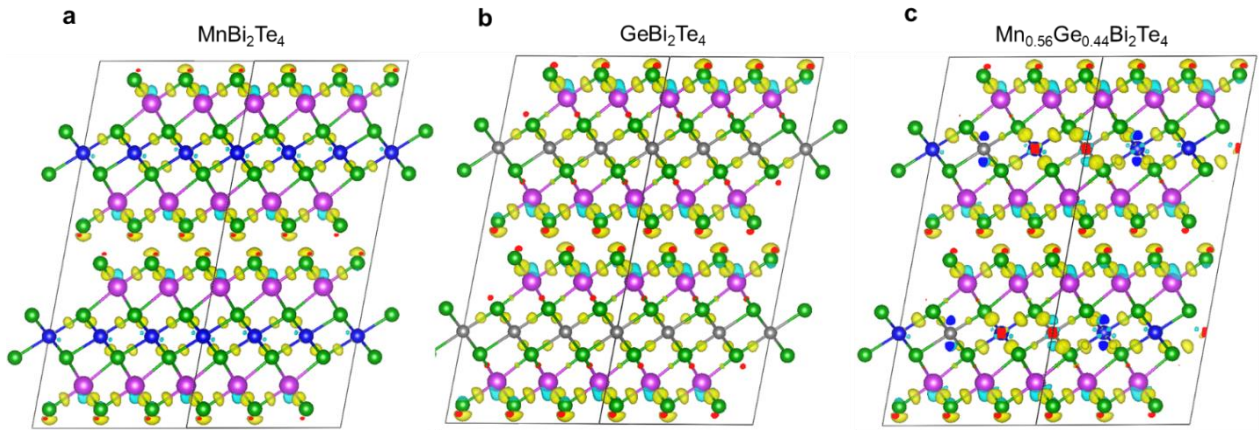

**Figure S6.** Charge difference after bonding for (a) $MnBi_2Te_4$, (b) $GeBi_2Te_4$ and (c) $Mn_{0.56}Ge_{0.44}Bi_2Te_4$.

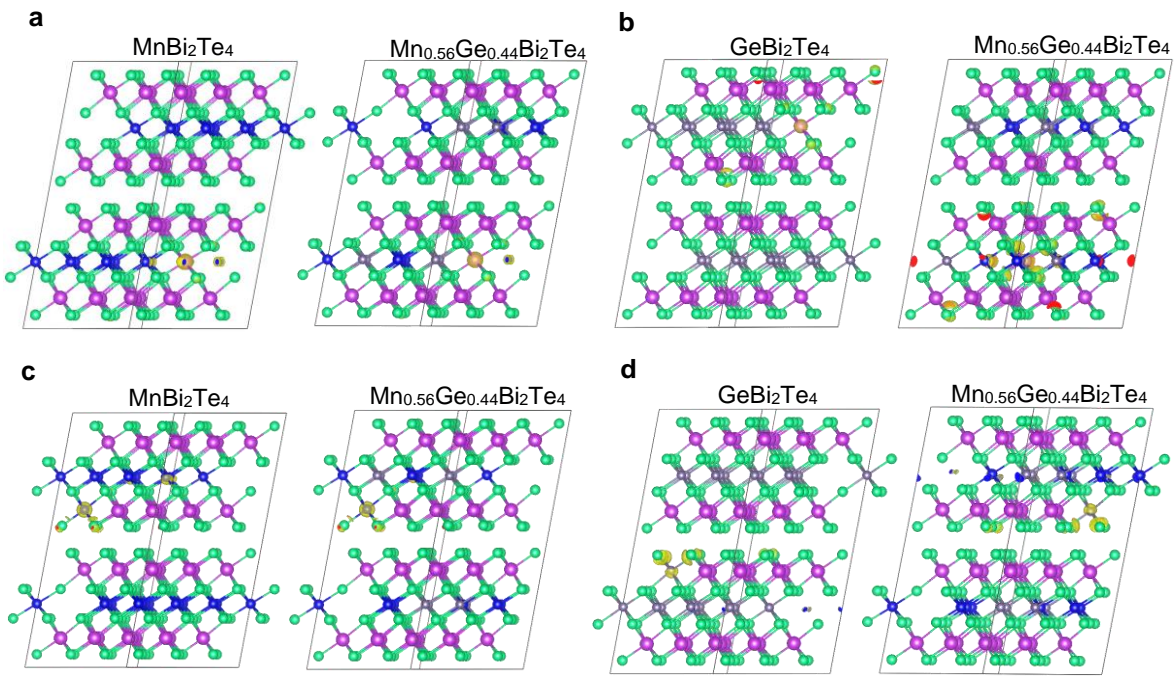

**Figure S7.** Charge difference distribution after the defect deformation for (a) $Bi_{Mn}$ antisite defects, (b) $Bi_{Ge}$ antisite defects, (c) $Mn_{Bi}$ antisite defects and (d) $Ge_{Bi}$ antisite defects.

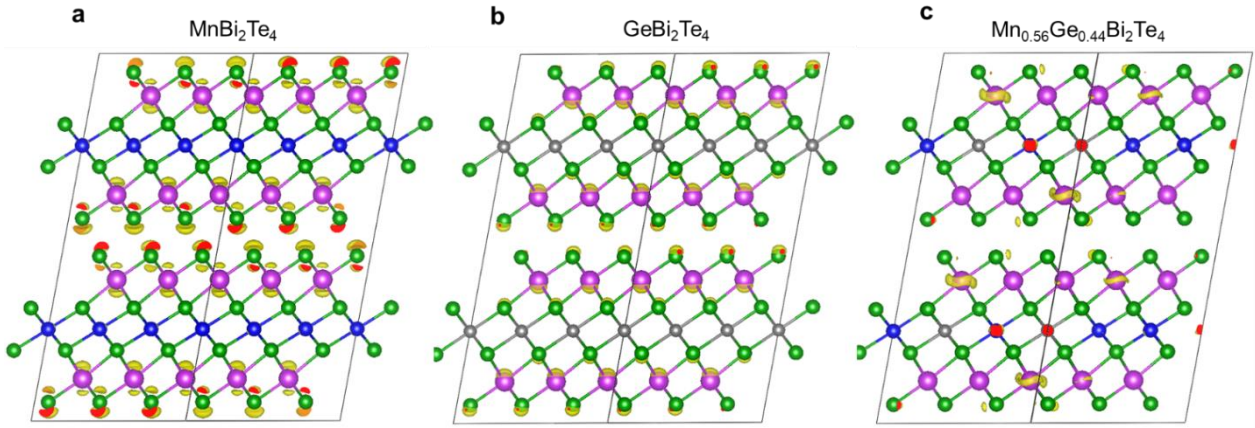

**Figure S8.** Partial charge distribution of VBM for (a) MnBi$_2$Te$_4$, (b) GeBi$_2$Te$_4$ and (c) Mn$_{0.56}$Ge$_{0.44}$Bi$_2$Te$_4$.

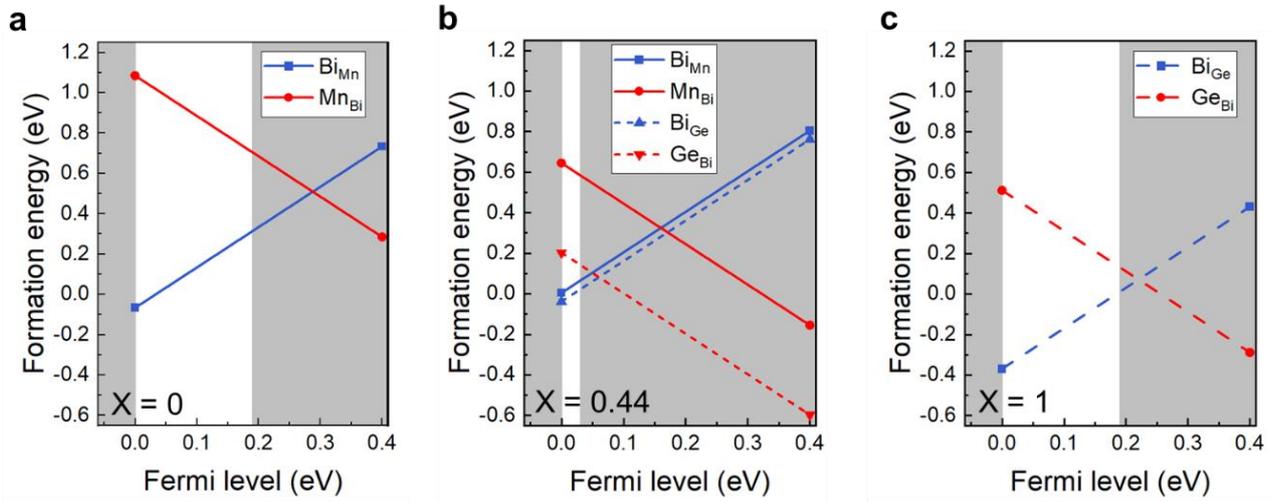

**Figure S9.** Calculated defect formation energies of (a) MBT ($x$ = 0), (b) MGBT ($x$ = 0.44) and (c) GBT ($x$ = 1), where two antisite defects are adopted on both septuple layers preserving the AFM orders.

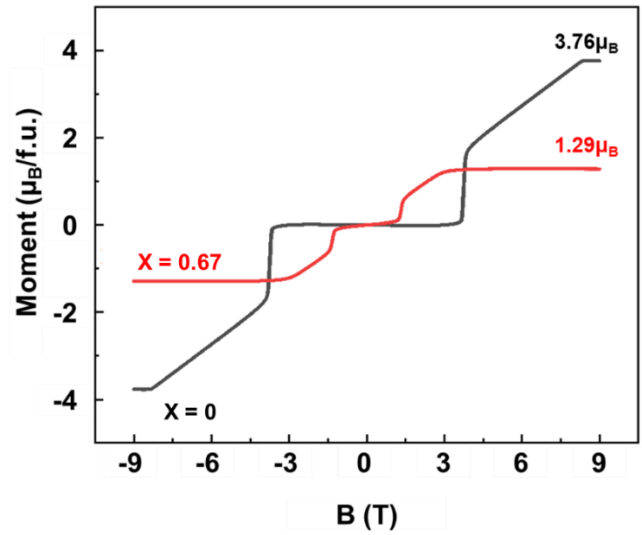

**Figure S10.** Magnetization versus magnetic field of MBT ($x = 0$) and MGBT ($x = 0.67$) without normalized.

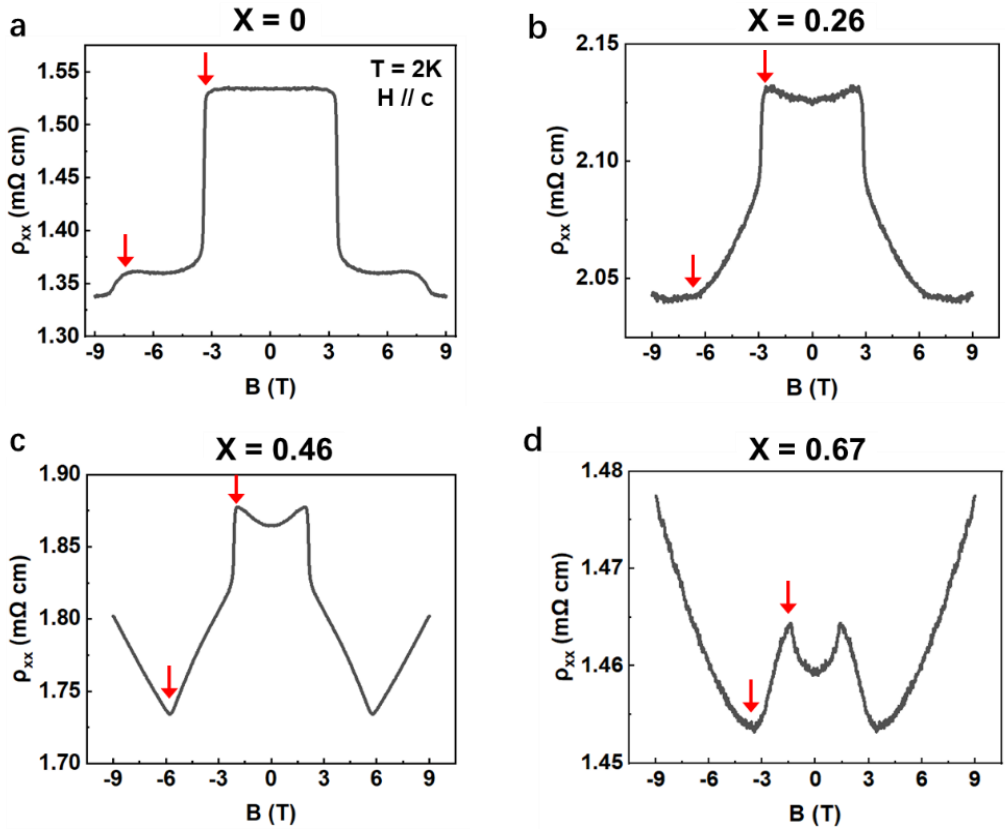

**Figure S11.** Magnetoresistance of $Mn_{1-x}Ge_xBi_2Te_4$ samples versus magnetic field with different doping levels of (a) $x = 0$, (b) $x = 0.26$, (c) $x = 0.46$ and (d) $x = 0.67$. The red arrows mark two transitions from AFM to CAFM and then to FM.

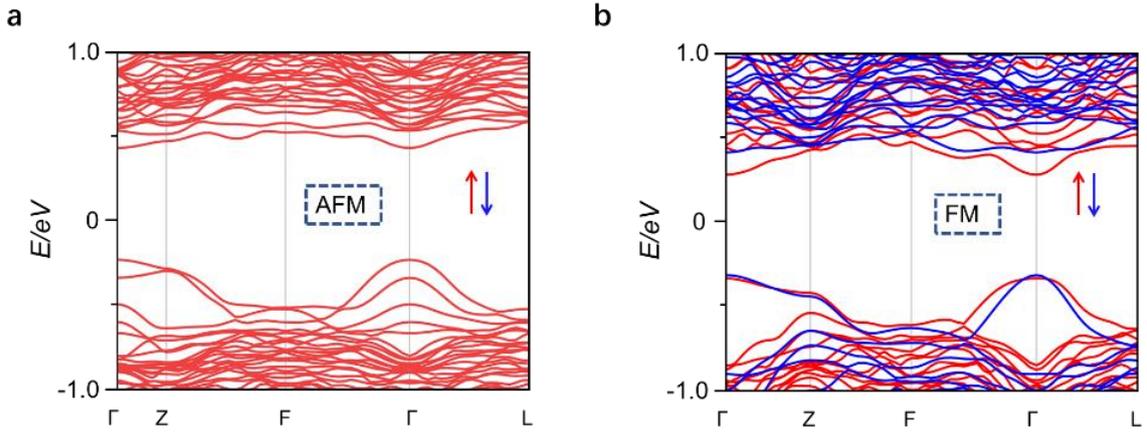

**Figure S12.** The calculated band structures of $Mn_{0.56}Ge_{0.44}Bi_2Te_4$ without SOC under the (a) AFM and (b) FM states. The red (blue) lines represent the spin-up (spin-down) bands.

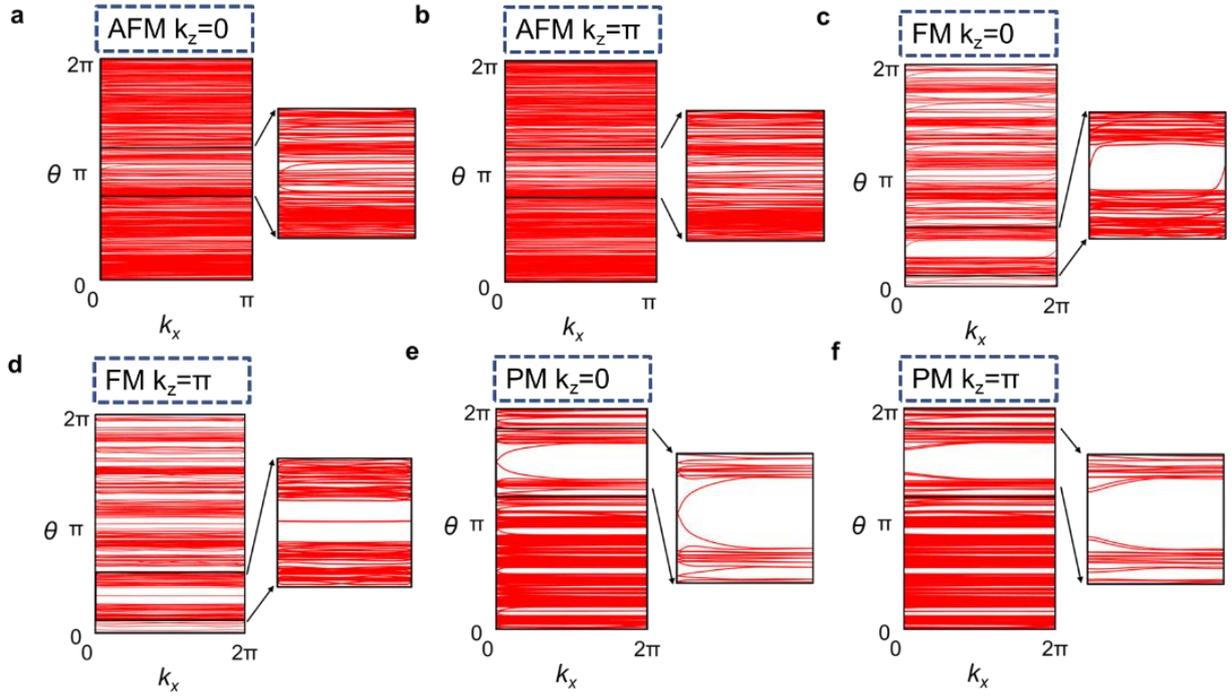

**Figure S13.** The evolution of WCC of $Mn_{0.56}Ge_{0.44}Bi_2Te_4$ under the (a) (b) AFM, (c) (d) FM, and (e) (f) PM states on the $k_z = 0$ and $k_z = \pi$ planes.

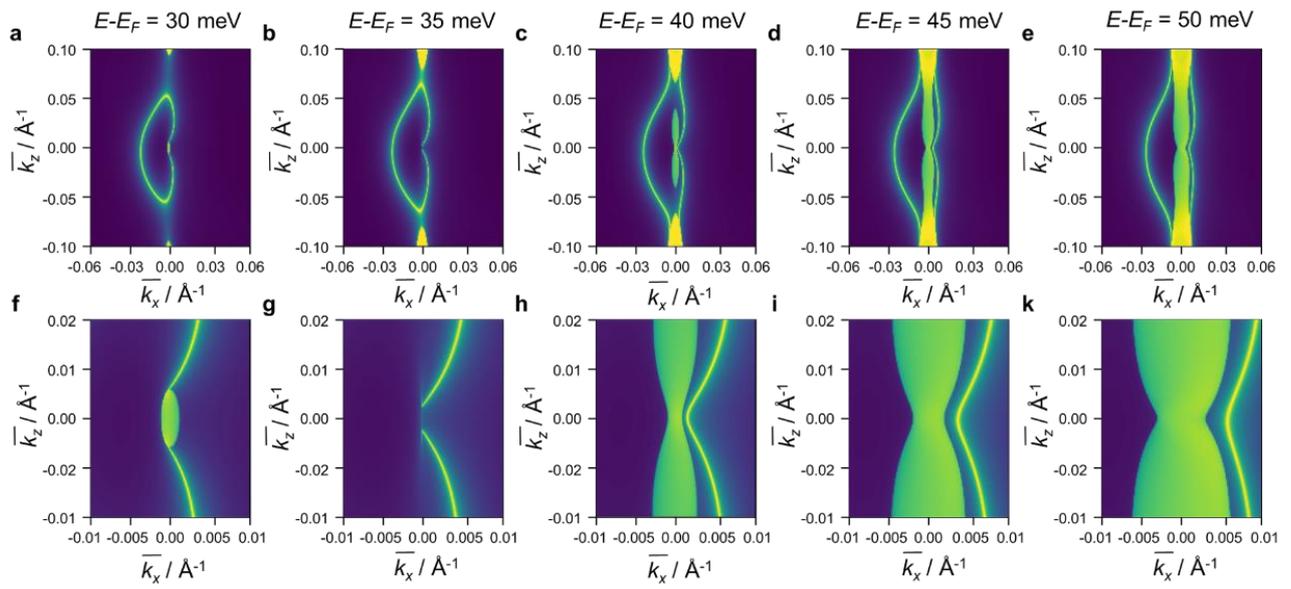

**Figure S14.** (a) - (e) Evolution of the surface Fermi arc for FM Weyl Mn$_{0.56}$Ge$_{0.44}$Bi$_2$Te$_4$ at different Fermi energies from 0.03 eV to 0.05 eV. (f) - (j) Enlarged views of the central area of (a) - (e).

# Supplementary Tables

**Table S1.** Summary of EDS results for Mn$_{1-x}$Ge$_x$Bi$_2$Te$_4$. The Te content is assumed to be a uniform value.

| Mn | Ge | Bi | Te |
|---|---|---|---|
| 0.955 | 0 | 2.132 | 4 |
| 0.703 | 0.253 | 2.124 | 4 |
| 0.537 | 0.479 | 2.084 | 4 |
| 0.255 | 0.671 | 2.103 | 4 |
| 0 | 0.932 | 2.170 | 4 |

**Table S2.** Energy of valence band maximum (VBM) for MnBi$_2$Te$_4$, GeBi$_2$Te$_4$ and Mn$_{0.56}$Ge$_{0.44}$Bi$_2$Te$_4$.

| System | E_VBM (eV) |
|---|---|
| MnBi$_2$Te$_4$ | 5.4874 |
| GeBi$_2$Te$_4$ | 5.2835 |
| Mn$_{0.56}$Ge$_{0.44}$Bi$_2$Te$_4$ | 5.5519 |